\documentclass[aps,prd,12pt,nofootinbib,superscriptaddress]{revtex4-2}
\usepackage{bm}
\usepackage{ulem}
\usepackage{soul}
\usepackage{comment}
\usepackage{amsmath}
\usepackage{amssymb}
\usepackage[mathscr]{euscript}
\usepackage{booktabs}
\usepackage{graphicx}
\usepackage{subfig}
\usepackage{float}
\usepackage{caption}
\usepackage{comment}
\usepackage{hyperref}
\usepackage{CJK}
\usepackage{color}
\usepackage{verbatim}
\hypersetup{colorlinks=true, linkcolor=red, citecolor=blue}

\newcommand{\nn}{\nonumber\\}

\begin{document}
\begin{CJK*}{UTF8}{gbsn}
\title{Constraint on gravitational-wave polarizations for space-based detectors with time-delay interferometry}

\author{Tong Jiang}
\email{jiangtong@hust.edu.cn}
\affiliation{College of Physics and Technology, Kunming University, Kunming, Yunnan 650214, China}

\author{Chunyu Zhang}
\email{Corresponding author.
chunyuzhang@yzu.edu.cn}
\affiliation{Center for Gravitation and Cosmology, College of Physical Science and Technology, Yangzhou University, Yangzhou 225009, China}

\begin{abstract}
Probing extra polarizations in gravitational waves (GWs) with space-based detectors is the most direct method for testing theories of gravity.
In this paper, by employing the second-generation time-delay interferometry (TDI) to cancel out the laser frequency noise in a rotating and flexing configuration with arm lengths varying linearly in time, we study the detectors' constraint ability on extra polarizations, and explore the impacts of TDI on the constraint of polarizations.
Working in the parametrized post-Einstein (ppE) waveform framework, we find that the constraints on extra polarizations are significantly weaker than those for the tensor mode.
For the tensor mode, the constraint ability for the detectors scales with signal-to-noise ratio (SNR).
At low frequency, due to signal cancellation effects, the SNR registered is lower for the detectors with TDI method than that with the simplified equal-arm Michelson interferometer method.
Therefore, tensor-mode constraints are degraded when TDI is applied.
Although the direct detection of extra polarizations remains challenging, the constraint ability of the space-based detectors on the vector mode is better than the scalar modes.
Besides, the detectability on extra polarizations will be enhanced with TDI, and the improvement of the constraint on the vector mode is less than scalar modes due to the inclination-dependent waveforms.
\end{abstract}

\maketitle
\end{CJK*}

\section{Introduction}
Since the advanced Laser Interferometer Gravitational-Wave Observatory (LIGO) Scientific Collaboration \cite{Harry:2010zz, LIGOScientific:2014pky} and Virgo Collaboration \cite{VIRGO:2014yos} detected the first gravitational wave (GW) event \cite{LIGOScientific:2016aoc} directly in 2015, we have entered the era of gravitational-wave astronomy.
Although Einstein's general relativity (GR) has passed a large number of experiments in the solar system and binary pulsars \cite{Stairs:2003eg,Will:2014kxa,Wex:2014nva}, the problems in the strong-field regime remain mostly unexplored \cite{Will:2005va}, and the theory of gravity is incomplete.
Therefore, by means of the GWs, it yields a new way to probe the evolution history of the Universe, study the properties of space-time in the strong-field and high-velocity regime, and further understand the nature of the gravity \cite{LIGOScientific:2016dsl,LIGOScientific:2016kwr,LIGOScientific:2016hpm,LIGOScientific:2021aug,Seto:2001qf,Kyutoku:2016zxn,eLISA:2013xep,PhysRevLett.121.129902,LIGOScientific:2018dkp,LIGOScientific:2019fpa,LIGOScientific:2020tif}.
Therefore, GWs provide a new way to probe the evolution history of the Universe, study the properties of space-time in the strong-field and high-velocity regime, and further understand the nature of the gravity.

In GR, GWs contain two massless tensor polarization modes propagating at the speed of light $c$, with an amplitude inversely proportional to the distance from the source.
However, in general metric theories of gravity, the number of polarization modes can reach up to six, including two tensor, two vector, and two scalar modes.
These additional polarization modes can exhibit distinct propagation velocities \cite{PhysRevD.8.3308,PhysRevD.95.104034,Hou:2017bqj,Gong:2017bru,Gong:2017kim,PhysRevD.97.084040,Gong:2018ybk,PhysRevD.98.104017,Hou:2018djz}, amplitude attenuation characteristics \cite{PhysRevD.106.044067,PhysRevD.97.104066,PhysRevD.97.104037,PhysRevD.85.024041}, and effective masses \cite{PhysRevD.57.2061,LIGOScientific:2021sio,Shoom:2022cmo}.
Therefore, using GWs to limit the polarization content \cite{LIGOScientific:2018dkp,LIGOScientific:2017ycc}, we can directly test the theories of gravity.

Although ground-based detectors have observed nearly a hundred GW events generated by compact binary coalescence so far, it has not been possible to use these signals to learn about the polarization content of GWs \cite{LIGOScientific:2016lio}.
Ground-based GW detectors can only detect the transient GW signals from merging sources, which last seconds to minutes in the frequency band around $1\sim 10^3 $ Hz.
The positional variation of detectors is nearly negligible during such signals.
Therefore, the polarization can not be detected with a single ground-based detector.
However, it is possible to determine or limit the polarization of GW by using a sufficiently large number of suitably oriented ground-based GW detectors.
The existing analysis of the GW data to constrain the amount of allowed non-GR polarizations is usually in an indirect and strongly model-dependent manner.
Such as measuring the orbital decay of the binary system is relative to the total radiation of the GW power \cite{Freire:2012mg,Stairs:2003eg}, the process can indeed constrain the power in extra polarizations, but does not provide the direct, model-independent information on the actual polarization content of the gravitational radiation.

In contrast, the proposed space-based GW detectors such as Laser Interferometer Space Antenna (LISA) \cite{Danzmann:1997hm,LISA:2017pwj}, Taiji \cite{Hu:2017mde}, and TianQin \cite{TianQin:2015yph} will detect GWs in a long-term running, signals of GWs which are from the short-period binary stars last for months to years in the lower frequency band around $10^{-4}\sim10^{-1}$ Hz.
During the mission of the space-based GW detector, the positional evolution of the detector in space can be treated as a continuous series of detections.
By taking a specific linear combination of the outputs in the network of detectors, it is possible to remove any tensorial signal present in the data \cite{PhysRevD.40.3884,PhysRevD.74.082005}.
It indicates that this configuration enables the validation of GW polarization by using a single space-based detector alone \cite{Zhang:2021fha}, and GW detection from inspiral to merger and ringdown with LISA, Taiji and TianQin will place more stringent constraints on alternative theories of gravity.
Therefore, it is important to study the capability of test polarizations for space-based GW detectors.
In this paper, we only consider the inspiral phase signals from the massive black hole binaries (MBHBs), which are one of the main targets of space-based detectors, neglecting the merger and ringdown phases for simplicity.

While using space-based detectors to constrain polarization contents, we need accurate waveform templates, but even in GR, the long evolution time of inspiral binaries makes the generation of accurate waveform templates for matched filtering a hard problem.
For the modified theory of gravity, the GW waveform templates including extra polarizations are highly model-dependent.
In most studies, one aims at the particular theories, or the effect that one wishes to constrain, then derives how GW would be deformed, if the GW is detected that is consistent with GR, a constraint on GW deformations can be placed, which in turn provides the constraint results for the specific parameters of polarizations.
The relative studies about constraints on the polarizations have been discussed extensively, please see \cite{Jiang:2021htl,Alsing:2011er,Yunes:2011aa,Yagi:2015oca,Will:2004xi,Yagi:2009zm,Berti:2013gfa,Sotani:2004rq,Healy:2011ef,Yunes:2025xwp,Quartin:2023tpl,Gao:2022hsn,Xie:2022wkx} and the reference therein.
However, to focus on the capability of space-based detectors to observe additional polarizations, a good solution is utilizing the model-independent waveform template to test the extent of response for different polarizations.
The parametrized post-Einstein (ppE) waveform \cite{Yunes:2009ke,Cornish:2011ys,Chatziioannou:2012rf} was introduced by Yunes and Pretorius, which is derived from the post-Newtonian (PN) approximation, to allow for the direct presence of additional polarization modes in the response function.
Most modified gravity theories, such as Brans-Dicke theory, massive gravity, and bimetric theory, can be characterized within the ppE framework, and it does offer a model-independent method for testing additional polarizations \cite{Shi:2022qno,Liu:2020mab,Chatziioannou:2012rf,Tahura:2018zuq,Yunes:2013dva,OBeirne:2019lwp,Liu:2020mab,Wu:2024yno,Wu:2024cal}.
Thus, in this paper, we use the ppE waveform as the waveform template to carry out a systematic study of the observing capability of polarizations with space-based detectors and their networks.

For the space-based GW detector, it cannot maintain exact equality between the arm lengths due to the orbital motion of satellites, so the laser frequency noise cannot be cancelled out completely.
To reduce the laser frequency noise, properly chosen time shifted and linearly combined data streams were proposed to synthesize virtual equal-arm interferometric measurements \cite{Tinto:1999yr,Tinto:1994kg,Armstrong_1999}.
This technique is known as time-delay interferometry (TDI) \cite{Armstrong_1999,Tinto:1999yr,Estabrook:2000ef,Tinto:2003vj,Vallisneri:2004bn}.
The first-generation TDI combinations can cancel out the laser frequency noise in a static unequal-arm configuration, and the second-generation TDI can cancel out the laser frequency noise in a rotating and flexing configuration with arm lengths varying
linearly in time \cite{Tinto:2003vj,Shaddock:2003dj,Cornish:2003tz}.
For more discussion on the TDI algorithm and its application to space-based GW detectors, please see \cite{Tinto:2004wu,Tinto:2014lxa,Tinto:2020fcc,Muratore:2020mdf} and references therein.
In this paper, considering the more realistic case, we take the second-generation TDI combinations in signal response and discuss the effects of constraints on the different gravitational-wave polarizations with space-based detectors.
The GW waveforms of different polarizations are derived in the ppE framework as the templates, and parameter analysis is given by the Fisher information matrix (FIM) method.

The paper is organized as follows. 
In Sec. \ref{meth}, we provide a brief overview of the GW waveforms of different polarizations, the signal response with the second-generation TDI combination, and the FIM method in parameter analysis.
In Sec. \ref{result}, we present the results and analyze the effects of constraint on the different gravitational-wave polarizations with different TDI combinations for the space-based detectors.
Finally, we summarize the results in Sec. \ref{discussion}.
In this paper, we use units with $G=c=1$, where $G$ is the gravitational constant, and $c$ is the speed of light.

\section{Methodology}\label{meth}
In this section, we give a brief overview of the methods for constraining GW polarizations with the space-based GW detectors, such as LISA, Taiji, and TianQin.
It mainly includes model-independent polarization waveforms in the ppE framework, the detector configurations of LISA, Taiji and TianQin, signal response with second-generation TDI combinations, and parameter analysis with the FIM method.

\subsection{GW waveforms of different polarization}
The GWs in many modified theories of gravity include all six polarizations, which are physical and must be accounted for in GW templates.
Therefore, a model-independent framework to test GR is developed: the ppE framework.
In this scheme, the enhanced GW templates of GR contain the additional parameters of modified theories, and the different values of these ppE parameters cause the enhanced templates to recover GR or predict from modified gravity theories.

For the GWs propagating along the direction $\mathbf{\hat{\Omega}}(\theta,\phi)$, we can define orthonormal coordinate system satisfying $\mathbf{\hat{\Omega}}=\hat{p}\times\hat{q}$, where $\hat{p}$ and $\hat{q}$ are perpendicular unit vectors.
The GW source is located at $\mathbf{\hat{w}}(\theta_s,\phi_s)=-\mathbf{\hat{\Omega}}$, where $\theta = \pi-\theta_s, \phi = \phi_s+\pi$.
We use $\psi$ to express the rotational degree of freedom along the GW propagating direction as the polarization angle, and the new basis vectors are
\begin{equation}
    \mathbf{\hat{m}}=\cos(\psi) \mathbf{\hat{p}}+\sin(\psi)\mathbf{\hat{q}},\qquad\mathbf{\hat{n}}=-\sin(\psi) \mathbf{\hat{p}}+\cos(\psi)\mathbf{\hat{q}}.
\end{equation}
Thus, we use these bases to derive six polarization tensors as follows
\begin{equation}
\begin{aligned}
\mathbf{\epsilon}^{+}_{ij}=\mathbf{\hat{m}}_i\mathbf{\hat{m}}_j-\mathbf{\hat{n}}_i\mathbf{\hat{n}}_j,  &\qquad  \mathbf{\epsilon}^{\times}_{ij}=\mathbf{\hat{m}}_i\mathbf{\hat{n}}_j+\mathbf{\hat{n}}_i\mathbf{\hat{m}}_j, \\
\mathbf{\epsilon}^{x}_{ij}=\mathbf{\hat{m}}_i\mathbf{\hat{\Omega}}_j+\mathbf{\hat{\Omega}}_i\mathbf{\hat{m}}_j, &\qquad   \mathbf{\epsilon}^{y}_{ij}=\mathbf{\hat{n}}_i\mathbf{\hat{\Omega}}_j+\mathbf{\hat{\Omega}}_i\mathbf{\hat{n}}_j, \\
\mathbf{\epsilon}^{b}_{ij}=\mathbf{\hat{m}}_i\mathbf{\hat{m}}_j+\mathbf{\hat{n}}_i\mathbf{\hat{n}}_j, &\qquad  \mathbf{\epsilon}^{l}_{ij}=\mathbf{\hat{\Omega}}_i\mathbf{\hat{\Omega}}_j.
\end{aligned}
\end{equation}
Therefore, the GW waveforms with entire polarizations can be expressed as
\begin{equation}\label{gbsig}
    h_{ij}(t)=\sum_{A}\mathbf{\epsilon}_{ij}^A h_A(t),
\end{equation}
where $A=\{+, \times, x, y, b, l\}$ denotes the different polarizations of GW.
The GW waveform in the improved ppE framework \cite{Chatziioannou:2012rf,Yunes:2009ke,Cornish:2011ys} can be written as 
\begin{equation}\label{eq:waveform}
    \begin{aligned}
    h_{+} & = \mathcal{A}_T \frac{1+\cos^{2}\iota}{2}\cos(2\Phi+2\Phi_0), \\
    h_{\times} & = \mathcal{A}_T \cos\iota\sin(2\Phi+2\Phi_0), \\
    h_{x} & = \mathcal{A}_V \cos\iota\cos(\Phi+\Phi_0), \\
    h_{y} & = \mathcal{A}_V \sin(\Phi+\Phi_0), \\
    h_{b} & = \mathcal{A}_B \sin\iota\cos(\Phi+\Phi_0), \\
    h_{l} & = \mathcal{A}_L \sin\iota\cos(\Phi+\Phi_0), \\
    \end{aligned}
\end{equation}
with
\begin{equation}\label{amp}
    \begin{aligned}
    \mathcal{A}_T&=\frac{4}{d_L} \mathcal{M}^{5/3}\omega^{2/3} , \\
    \mathcal{A}_V&= \frac{\alpha_V}{d_L}\mathcal{M}^{4/3}\omega^{1/3}, \\
    \mathcal{A}_B&= \frac{\alpha_B}{d_L}\mathcal{M}^{4/3}\omega^{1/3}, \\
    \mathcal{A}_L&= \frac{\alpha_L}{d_L}\mathcal{M}^{4/3}\omega^{1/3}, \\
    \end{aligned}
\end{equation}
where $\iota$ is the inclination angle, $\Phi$ is the orbital phase, $\omega$ is orbital angular frequency, $\Phi_0$ is initial phase of orbit, $\alpha_{V, B, L}$ is dimentionless ppE parameters of different GW polarizations, $d_L$ is luminosity distance, $\mathcal{M}=(m_1m_2)^{3/5}/(m_1+m_2)^{1/5}$ is the chirp mass, $m_1$ and $m_2$ are masses of binary black holes.
Using the balance law to relate the rate of change of binding energy to the GW luminosity, the evolution of orbital angular frequency in the ppE framework can be expressed as
\begin{equation}\label{evoangular}
    \frac{\mathrm{d}\omega}{\mathrm{d}t}=\alpha_D\eta^{2/5}\mathcal{M}\omega^3+\alpha_Q\mathcal{M}^{5/3}\omega^{11/3} ,
\end{equation}
where the $\alpha_D$ and $\alpha_Q$ are the ppE parameters which describe the orbital angular frequency contributions of dipole and quadrupole radiation, and $\eta=m_1m_2/(m_1+m_2)^2$ is the symmetric mass ratio.
The evolution of orbital angular frequency $\omega(t)$ can be derived by integrate the Eq. \eqref{evoangular}, and further get orbital phase $\Phi(t)$ with $\Phi=\int \omega(t)\ dt$ as

\begin{equation}\label{omegat}
\begin{split}
t-t_0= & \frac{3 \alpha_Q \mathcal{M}^{-1 / 3}}{4 \eta^{4 / 5} \alpha_D^2}\left(\frac{1}{\omega^{4 / 3}}-\frac{1}{\omega_0^{4 / 3}}\right)-\frac{\left(\omega^{-2}-\omega_0^{-2}\right)}{2 \mathcal{M} \eta^{2 / 5} \alpha_D}-\frac{3 \alpha_Q^2 \mathcal{M}^{1 / 3}}{2 \eta^{6 / 5} \alpha_D^3}\left(\frac{1}{\omega^{2 / 3}}-\frac{1}{\omega_0^{2 / 3}}\right)\\
& +\frac{3 \mathcal{M} \alpha_Q^3}{2 \eta^{8 / 5} \alpha_D^4} \log \left(\frac{\eta^{2 / 5} \omega^{-2 / 3} \alpha_D+\mathcal{M}^{2 / 3} \alpha_Q}{\eta^{2 / 5} \omega_0^{-2 / 3} \alpha_D+\mathcal{M}^{2 / 3} \alpha_Q}\right)\\
= &- \frac{3}{8 \mathcal{M}^{5/3} \alpha_Q} \left(\frac{1}{\omega^{8/3}} - \frac{1}{\omega_0^{8/3}}\right)+ \frac{3 \alpha_D \eta^{2/5}}{10 \mathcal{M}^{7/3} \alpha_Q^2} \left(\frac{1}{\omega^{10/3}} - \frac{1}{\omega_0^{10/3}}\right) \\
&-\frac{\alpha_D^2 \eta^{4/5}}{4 \mathcal{M}^3 \alpha_Q^3} \left(\frac{1}{\omega^4} - \frac{1}{\omega_0^4}\right) + O(\alpha_D^3),
\end{split}
\end{equation}
\begin{equation}\label{phit}
\begin{split}
\Phi-\Phi_0=&\frac{3 \alpha_Q\mathcal{M}^{-1/3}}{\alpha_D^2 \eta ^{4/5}}\left(\frac{1}{\omega^{1/3}}-\frac{1}{\omega_0^{1/3}}\right)-\frac{(\omega^{-1}-\omega_0^{-1})}{\alpha_D \eta ^{2/5}  \mathcal{M}}\\
&+\frac{3 \alpha_Q^{3/2}}{\alpha_D^{5/2} \eta }\left(\tan ^{-1}\left(\frac{\sqrt{\alpha_Q} (\mathcal{M} \omega)^{1/3}}{\sqrt{\alpha_D} \eta^{1/5}}\right)-\tan ^{-1}\left(\frac{\sqrt{\alpha_Q} (\mathcal{M} \omega_0)^{1/3}}{\sqrt{\alpha_D}\eta^{1/5}}\right)\right)\\
= &-\frac{3}{5 \mathcal{M}^{5/3} \alpha_Q} \left(\frac{1}{\omega^{5/3}} - \frac{1}{\omega_0^{5/3}}\right) + \frac{3\alpha_D  \eta^{2/5}}{7 \mathcal{M}^{7/3} \alpha_Q^{2}} \left(\frac{1}{\omega^{7/3}} - \frac{1}{\omega_0^{7/3}}\right)\\
&- \frac{\alpha_D^2 \eta^{4/5}}{3 \mathcal{M}^{3} \alpha_Q^{3}} \left(\frac{1}{\omega^{3}} - \frac{1}{\omega_0^{3}}\right)+ O(\alpha_D^3),
\end{split}
\end{equation}
where $t_0$ and $\omega_0$ are the initial condition of time and angular frequency.
Since $\alpha_D$ is a small quantity, we use Taylor expansion in our calculations to accelerate computation and improve accuracy.

Note we expect to get the evolution of $\omega(t)$ and $\Phi(t)$ with $t$ from Eq. \eqref{omegat} and \eqref{phit}, but the strict solutions are impossible.
Here we solve the equations point by point numerically, and finally provide the results of $\omega(t)$ and $\Phi(t)$.
Substitute the results into Eqs. \eqref{gbsig} - \eqref{amp} to derive the GW waveforms with complete polarizations.

\subsection{The configurations of space-based detectors}
The proposed space-based detectors have two different constellations.
For LISA/Taiji, the spacecraft is in the heliocentric orbit behind/ahead of the Earth by about $20^\circ$,
the inclination angle between the plane of constellations and ecliptic plane is $60^\circ$, and they keep the geometry of an almost equilateral triangle.
To the first order of orbital eccentricity, the coordinates $\textbf{x}(t)$ of the three spacecrafts in the heliocentric reference frame are given by \cite{Rubbo:2003ap,Cutler:1997ta}
\begin{eqnarray}
x(t) &=& R \cos \alpha+\frac{1}{2} e R\left[\cos (2 \alpha-\beta)-3 \cos \beta\right] , \nonumber\\
y(t) &=& R \sin \alpha+\frac{1}{2} e R\left[\sin (2 \alpha-\beta)-3 \sin \beta\right] , \nonumber\\
z(t) &=& -\sqrt{3} \,e R \cos (\alpha-\beta) ,
\end{eqnarray}
where $R=1$ AU, $e=L/(2\sqrt{3}R)$ is the orbital eccentricity, the average arm length $L$ is $2.5\times10^9$ m$/3\times10^9$ m for LISA /Taiji, $\alpha=2\pi f_m t+\kappa$ with $f_m=1/$year, and $\beta=2\pi n/3+\lambda $ $ (n=0,1,2)$.
Here $\kappa$ and $\lambda$ are the initial ecliptic longitude and orientation of the spacecraft, respectively; $\kappa=-20^\circ$ for Taiji and $\kappa=20^\circ$ for LISA.

For TianQin, the spacecrafts are in the geocentric orbit around the Earth and rotates around the Sun, the arm length is $L=1.73\times10^8$ m, and the normal vector of the detector plane points to the source RX J0806.3+1527 at $(\theta_{tq}, \phi_{tq})=(-4.7^{\circ}, 120.5^{\circ})$.
The $n$-th spacecraft's position vector $\textbf{x}(t)$ to the first order of orbital eccentricity in the heliocentric-ecliptic coordinate system is given by \cite{Hu:2018yqb}
\begin{eqnarray}
  x_n(t)&&=R_1\big(\cos\phi_{tq}\sin\theta_{tq}\sin(\alpha_n-\beta')+\cos(\alpha_n-\beta')\sin\phi_{tq}\big)\nonumber\\
  &&+R\cos(\alpha-\beta)+\frac{Re}{2}\big(\cos2(\alpha-\beta)-3\big) ,\nonumber\\ 
  y_n(t)&&=R_1\big(\sin\phi_{tq}\sin\theta_{tq}\sin(\alpha_n-\beta')-\cos(\alpha_n-\beta')\cos\phi_{tq}\big)\nonumber\\
  &&+R\sin(\alpha-\beta)+\frac{Re}{2}\sin2(\alpha-\beta) ,\nonumber\\
  z_n(t)&&=-R_1\sin(\alpha_n-\beta')\cos\theta_{tq},\end{eqnarray}
where $R=1$ AU, $R_1=1\times10^8$ m is the semi-major axis of the spacecraft orbit, $e=0.0167$ is the eccentricity of the geocenter orbit around the Sun.
Here $\alpha (t) = 2\pi f_{\rm{m}}t + \kappa_0$ is the mean ecliptic longitude 
of the geocenter in the heliocentric-ecliptic coordinate system, $f_{\rm{m}}=1/\rm{year}$ is the modulation frequency due to the orbital motion around the Sun, 
and $\kappa_0$ is the mean ecliptic longitude measured from 
the vernal equinox at $t=0$. $\beta$ is the longitude of the perihelion. 
The spacecraft orbit phase $\alpha _n(t) = 2\pi {f_{\rm{sc}}}t +  \frac{2\pi}{3}(n - 1) + \lambda$, $\lambda$ is the initial orbit phase of the first $(n=1)$ spacecraft, $f_{\rm{sc}} \approx 1/(3.65~\rm{day})$ is the modulation frequency due to the rotation of the detector around the guiding center. $\beta'$ is the angle to the perigee of the spacecraft orbit.
We set $\kappa_0=\beta=\lambda=\beta'=0$ in the calculation.

\subsection{Equal-arm Michelson interferometer}

We simultaneously use both the simplified equal-arm Michelson interferometer method and the second-generation TDI method employed in actual detection to provide GW polarization constraint results for comparison.

The configurations of space-based GW detectors are generally equilateral triangles.
We can model every detector of this kind as a combination of two independent LIGO-like detectors (``I" and ``II") with the opening angle $\gamma=\pi/3$.

The response of the one-way transmission in the arm of the interferometric GW detector is defined as
\begin{equation}
s(f)=\sum_A F^A h_A(f)e^{i\Phi_D},
\end{equation}
where 
\begin{equation}
F^A=D^{ij} e^A_{ij}
\end{equation}
is the angular response function $F^A$  for polarization $A$,
\begin{equation}
D^{ij}=\frac{1}{2}[\hat{u}^i \hat{u}^j T(f,\hat{u}\cdot\mathbf{\hat{\Omega}})-\hat{v}^i \hat{v}^j  T(f,\hat{v}\cdot\mathbf{\hat{\Omega}})]
\end{equation}
is the detector tensor $D^{ij}$,
$\hat{u}$ and $\hat{v}$ are the unit vectors along the arms of the detector,
\begin{equation}
\label{transferfunction}
\begin{split}
T(f,\hat{u}\cdot\mathbf{\hat{\Omega}})=\frac{1}{2}&\left\{\text{sinc}\left[\frac{f}{2f^*}(1-\hat{u}\cdot\mathbf{\hat{\Omega}})\right]\exp\left[-i\frac{f}{2f^*}(3+\hat{u}\cdot\mathbf{\hat{\Omega}})\right] \right.\\
& +\left. \text{sinc}\left[\frac{f}{2f^*}(1+\hat{u}\cdot\mathbf{\hat{\Omega}})\right]\exp\left[-i\frac{f}{2f^*}(1+\hat{u}\cdot\mathbf{\hat{\Omega}})\right]\right\}
\end{split}
\end{equation}
is the transfer function,
$\text{sinc}(x)=\sin x/x$, $f^*=1/(2\pi L)$ is the transfer frequency of the detector,
and $L$ is the arm length of the detector.
Note that in the low frequency limit $f\ll f^*$, we have $T(f,\hat{u}\cdot\hat{w})\rightarrow1$.

Due to the cancellation of frequency noises in the equal-arm interferometry, the dominant detector noises are the acceleration noise and the single-link optical metrology noise.
The noise curve is given by \cite{Cornish:2018dyw}
\begin{equation}\label{EqPn}
	P_n(f) = \frac{S_x}{L^2}  + \frac{2[1+\cos^2(2\pi fL)]S_a}{(2\pi f)^4 L^2},
\end{equation}
where $S_x$ is the position noise, and $S_a$ is the acceleration noise, $L$ is the arm length.
For LISA, $S_x = (1.5 \times 10^{-11} \ {\rm m})^2  \ {\rm Hz}^{-1}$, $S_a = (3 \times 10^{-15} \ {\rm m}\, {\rm s}^{-2})^2 \left[1+\left(\frac{0.4 \text{ mHz}}{f}\right)^2\right]\ {\rm Hz}^{-1}$, and $L=2.5\times10^9$ m \cite{Audley:2017drz}.
For Taiji, $S_x = (8\times10^{-12} \ {\rm m})^2  \ {\rm Hz}^{-1}$, $S_a = (3\times10^{-15} \ {\rm m}\, {\rm s}^{-2})^2\ {\rm Hz}^{-1}$, and $L=3\times10^9$ m \cite{Ruan:2020smc}.
For TianQin, $S_x = (10^{-12}\ {\rm m})^2  \ {\rm Hz}^{-1}$, $S_a = (10^{-15} \ {\rm m}\, {\rm s}^{-2})^2\  {\rm Hz}^{-1}$, and $L=\sqrt{3}\times10^8$ m \cite{Luo:2015ght}.

\subsection{Time-delay interferometry and signal response}
Due to the effect of arm lengths of detectors varying with time, to deal with the laser frequency noise in the rotating and flexing configuration, the second-generation TDI method should be considered.

To consider the TDI method, although, there are numerous TDI combinations which can be derived from the four generators $(\alpha, \beta, \gamma, \zeta)$ \cite{Prince:2002hp}, we focus on the so-called optimal combination $(\mathcal{A},\mathcal{E}, \mathcal{T})$, which can be obtained by linearly combining the $(X, Y, Z)$ channels of the Michelson configuration, in this paper.
For the second-generation $(X, Y, Z)$ combination, the signal response are 
\begin{align}
s_X = &\ z_{13}+\mathcal{D}_{13} z_{31}+\mathcal{D}_{131} z_{12}+\mathcal{D}_{1312} z_{21}+\mathcal{D}_{13121} z_{12}+\mathcal{D}_{131212} z_{21}\nn
&+\mathcal{D}_{1312121} z_{13}  +\mathcal{D}_{13121213} z_{31}\nn
&-\Big[z_{12}+\mathcal{D}_{12} z_{21}+\mathcal{D}_{121} z_{13}+\mathcal{D}_{1213} z_{31}+\mathcal{D}_{12131} z_{13}+\mathcal{D}_{121313} z_{31} \nn
& +\mathcal{D}_{1213131} z_{12}+\mathcal{D}_{12131312} z_{21}\Big], \\
s_Y = &\ z_{21}+\mathcal{D}_{21} z_{12}+\mathcal{D}_{212} z_{23}+\mathcal{D}_{2123} z_{32}+\mathcal{D}_{21232} z_{23}+\mathcal{D}_{212323} z_{32}\nn
&+\mathcal{D}_{2123232} z_{21}  +\mathcal{D}_{21232321} z_{12}\nn
&-\Big[z_{23}+\mathcal{D}_{23} z_{32}+\mathcal{D}_{232} z_{21}+\mathcal{D}_{2321} z_{12}+\mathcal{D}_{23212} z_{21}+\mathcal{D}_{232121} z_{12} \nn
& +\mathcal{D}_{2321212} z_{23}+\mathcal{D}_{23212123} z_{32}\Big], \\
s_Z = &\ z_{32}+\mathcal{D}_{32} z_{23}+\mathcal{D}_{323} z_{31}+\mathcal{D}_{3231} z_{13}+\mathcal{D}_{32313} z_{31}+\mathcal{D}_{323131} z_{13}\nn
&+\mathcal{D}_{3231313} z_{32}  +\mathcal{D}_{32313132} z_{23}\nn
&-\Big[z_{31}+\mathcal{D}_{31} z_{13}+\mathcal{D}_{313} z_{32}+\mathcal{D}_{3132} z_{23}+\mathcal{D}_{31323} z_{32}+\mathcal{D}_{313232} z_{23} \nn
& +\mathcal{D}_{3132323} z_{31}+\mathcal{D}_{31323231} z_{13}\Big],
\end{align}
where $z_{ij}$ is the phase difference time series measured at reception at spacecraft $i$ with transmission from spacecraft $j$ along $L_{ij}$, and the $\mathcal{D}$ is the delay operator, are given by
\begin{align}
z_{ij}(t)= \frac{\mathbf{\hat{u}}_{ij}(t)\cdot\mathbf{\epsilon}^{A}\cdot\mathbf{\hat{u}}_{ij}(t)}{2(1-\mathbf{\hat{u}}_{ij}(t)\cdot \mathbf{\hat{\Omega}})} &\Big[h_{A}(t- L_{ij}(t)- \mathbf{x}_{j}(t)\cdot\mathbf{\hat{\Omega}})-h_{A}(t- \mathbf{x}_{i}(t)\cdot\mathbf{\hat{\Omega}})\Big],\\
\mathcal{D}_{i_{1} i_{2} \ldots i_{n}} z_{ij}(t) &= z_{ij}\left(t-\sum_{k=1}^{n-1} L_{i_{k} i_{k+1}}(t)\right),
\end{align}
$L_{ij}(t)$ is the delay time along link $ij$ at reception time $t$, $\mathbf{x}_{i}(t)$ represents the position of the spacecraft $i$ at time $t$, and the $\mathbf{\hat{u}}_{ij}$ is unit vector pointing from the spacecraft $j$ to the spacecraft $i$, are given by
\begin{align}
L_{ij}(t)&\simeq|\mathbf{x}_{i}(t) - \mathbf{x}_{j}(t)|,\nonumber\\
\mathbf{\hat{u}}_{ij}(t) &= \frac{\mathbf{x}_{i}(t) - \mathbf{x}_{j}(t)}{|\mathbf{x}_{i}(t) - \mathbf{x}_{j}(t)|}.\nonumber
\end{align}
For example
$\mathcal{D}_{13121} z_{12}(t)=z_{12}\left(t- L_{13}(t)- L_{31}(t)- L_{12}(t)- L_{21}(t)\right)$.

After linearly combining the Michelson configuration, the $(\mathcal{A},\mathcal{E}, \mathcal{T})$ combination can be derived as
\begin{align}
s_\mathcal{A} =& \frac{1}{\sqrt{2}}(s_Z-s_X) \nn
s_\mathcal{E} =& \frac{1}{\sqrt{6}}(s_X-2 s_Y+s_Z) \nn
s_\mathcal{T} =& \frac{1}{\sqrt{3}}(s_X+s_Y+s_Z)
\end{align}

The noise power spectral densities of the optimal combination $(\mathcal{A},\mathcal{E}, \mathcal{T})$ are
\begin{align}
P^\mathcal{A}_n(f) =& P^\mathcal{E}_n(f)\nonumber \\
=&8 \sin ^2(2 \pi fL)\bigg[2 S_{y}^{\text {pm }}\bigg(3+2 \cos (2 \pi fL)+\cos (4 \pi fL)\bigg)+S_{y}^{\text {op}}\bigg(2+\cos (2 \pi fL)\bigg)\bigg], \nn
P^\mathcal{T}_n(f) =& 16 S_{y}^{\text {op}}\bigg(1-\cos (2 \pi fL)\bigg) \sin ^2(2 \pi fL)  +128 S_{y}^{\text {pm}} \sin ^2(2 \pi fL) \sin ^4(\pi fL),
\end{align}
where $S_{y}^{\text {pm}} = S_a/(2\pi f)^2$ and $S_{y}^{\text {op}} = S_x\left(2\pi f\right)^2$ are the proof mass noise and optical path noise expressed as fractional frequency fluctuation spectra, respectively.
For LISA,
$S_{y}^{\text {pm}}(f)= 2.5 \times 10^{-48} \left(\frac{1 \text{ Hz}}{f}\right)^{2}
\left[1+\left(\frac{0.4 \text{ mHz}}{f}\right)^2\right]  \text{ Hz}^{-1}$, $S_{y}^{\text {op}}(f) = 9.9 \times 10^{-38} \left(\frac{f}{1 \text{ Hz}}\right)^{2} \text{ Hz}^{-1}$;
For Taiji, $S_{y}^{\text {pm}}(f)= 2.5 \times 10^{-48}\left(\frac{1 \text{ Hz}}{f}\right)^{2} \text{ Hz}^{-1}$,
$S_{y}^{\text {op}}(f) =2.8 \times 10^{-38} \left(\frac{f}{1 \text{ Hz}}\right)^{2} \text{ Hz}^{-1}$;
For TianQin, $S_{y}^{\text {pm}}(f)= 2.8 \times 10^{-49} \left(\frac{1 \text{ Hz}}{f}\right)^{2} \text{ Hz}^{-1}$,
$S_{y}^{\text {op}}(f) = 4.4 \times 10^{-40} \left(\frac{f}{1 \text{ Hz}}\right)^{2} \text{ Hz}^{-1}$.

\subsection{Signal analysis}
The time-domain signal response of second-generation TDI for $(\mathcal{A},\mathcal{E}, \mathcal{T})$ channels are derived above.
Applying the Fourier transform numerically, we can derive the frequency-domain response and utilize the FIM method to estimate the parameter errors.
For the two frequency-domain signals $s^1(f)$ and $s^2(f)$, the inner product $\langle s^1|s^2\rangle$ is defined as
\begin{equation}\label{overlap}
    \langle s^1|s^2\rangle=2\sum_{\alpha=\mathcal{A}, \mathcal{E}, \mathcal{T}} \int_0^{+\infty } \frac{s^1_\alpha(f) s_{\alpha}^{2\,*}(f)+ s_\alpha^2(f) s_\alpha^{1\,*}(f)}{ P^\alpha_n(f)}\,df,
\end{equation}
where the star index denotes the complex conjugate of the signal.
The SNR of GW is 
\begin{equation}
\label{snr}
\begin{split}
\rho^2=& \sum_{\alpha=\mathcal{A}, \mathcal{E}, \mathcal{T}} \left< s_\alpha | s_\alpha \right>\\
=&4\sum_{\alpha=\mathcal{A}, \mathcal{E}, \mathcal{T}}  \int_{0}^{\infty}\frac{ s_\alpha(f) s_\alpha^*(f)}{P_{n}^\alpha(f)}df.
\end{split}
\end{equation}
To estimate the uncertainty of ppE parameters, we apply the FIM method.
The FIM is defined as
\begin{equation}
\label{tgamma}
\begin{split}
\Gamma_{ij}=& \sum_{\alpha=\mathcal{A}, \mathcal{E}, \mathcal{T}}\left<\frac{\partial s_\alpha}{\partial \bm{\Lambda}_i}\left|\frac{\partial s_\alpha^*}{\partial \bm{\Lambda}_j}\right.\right>\\
=&4\ \text{Re} \int_{0}^{\infty}\frac{\partial_i s_\alpha(f)\partial_j s_\alpha^*(f)}{P_{n}^\alpha(f)}df\\
\end{split}
\end{equation}
where $\alpha$ denotes the TDI combination, $\partial_i s_\alpha=\partial s_\alpha/\partial \bm{\Lambda}_i$ and $\bm{\Lambda}_i$ is the $i$th parameter of GW signals.
The covariance matrix $\sigma_{ij}$ between the parameters errors $\Delta\bm{\Lambda}_i=\bm{\Lambda}_i-\langle\bm{\Lambda}_i\rangle$ and $\Delta\bm{\Lambda}_j$ can be approximated by the inverse of the Fisher matrix in the large SNR limit,
\begin{equation}
\sigma_{ij}=\left\langle\Delta\bm{\Lambda}_i\Delta\bm{\Lambda}_j\right\rangle\approx (\Gamma^{-1})_{ij}.
\end{equation}

\section{Results}\label{result}
In this section, we show the results of the constraint on polarization for space-based detectors with the second-generation TDI method, and the impact of TDI on the constraint of polarization.

To study the effect of constraint on polarization with different space-based GW detectors and their network, taking into account the second-generation TDI configuration, the $(\mathcal{A},\mathcal{E}, \mathcal{T})$ combination as signal response, we focus on the parameter estimation of the five ppE parameters $(\alpha_D,\alpha_Q, \alpha_V, \alpha_L, \alpha_B)$ with fiducial values of $\alpha_D=0$, $\alpha_Q=96/5$, $\alpha_V=0$, $\alpha_L=0$, $\alpha_B=0$.
We choose 3 equal-mass MBHBs with total mass $M=(2\times10^4,~2\times10^5,~2\times10^6)~M_\odot$ as the GW sources at redshift $z=1$.
We do not consider the MBHB with total mass $M=2\times10^7~M_\odot$ because the signal-to-noise ratio (SNR) for its inspiral signal is too low.
The corresponding luminosity distance can be calculated in the spatially-flat $\Lambda$CDM Universise with the Planck 2018 results \cite{Planck:2018vyg}: the Hubble constant $H_0=67.66\ \text{km}/(\text{s}\cdot\text{Mpc})^{-1}$, the cosmological constant $\Omega_\Lambda=0.6889$, the matter energy density parameter $\Omega_{\text{m}}=0.3111$.
We simulate 100 GW sources for different MBHB with randomly sampled parameters $(\theta_s,\phi_s, \psi,\iota)$, 
where $\cos\theta_s \sim U\left(-1, 1\right)$, $\phi_s\sim U\left(-\pi, \pi\right)$, $\psi_s\sim U\left(0, \pi\right)$ and $\cos\iota \sim U\left(-1, 1\right)$.
We only consider the GW signals of the inspiral phase, and 60 days observation before the binaries reach to the innermost stable circular orbit (ISCO).
For the MBHBs, the upper cutoff frequency is the ISCO frequency $(6^{3/2}\pi M)^{-1}$, and the value of the lower frequency can be derived from Eq.\eqref{omegat}.
The estimation errors of ppE parameters are derived by FIM, and we take the mean value of 100 GW sources as the results to compare the effect of constraint on polarization with LISA, Taiji, TianQin, and their network.

\begin{table}
    \tabcolsep=1.5cm
    \renewcommand\arraystretch{0.5}
    \begin{tabular}{c | c}\hline\hline
\scriptsize{$M=2\times10^4M_\odot$}&\scriptsize{\text{TDI}}\\ \hline
 \scriptsize{\text{LISA}}&\scriptsize{($2.1 \times 10^{-6}$, $1.3 \times 10^{-4}$, $6.2 \times 10^{-3}$, $6.0 \times 10^{-2}$, $7.5 \times 10^{-2}$)} \\ \hline
 \scriptsize{\text{Taiji}}&\scriptsize{($1.1 \times 10^{-6}$, $6.8 \times 10^{-5}$, $1.4 \times 10^{-3}$, $4.8 \times 10^{-3}$, $7.6 \times 10^{-3}$)} \\ \hline
 \scriptsize{\text{TianQin}}&\scriptsize{($6.9 \times 10^{-5}$, $3.4 \times 10^{-3}$, $7.1 \times 10^{-2}$, $4.8 \times 10^{0}$, $4.8 \times 10^{0}$)} \\ \hline
 \scriptsize{\text{network}}&\scriptsize{($9.7 \times 10^{-7}$, $5.9 \times 10^{-5}$, $1.3 \times 10^{-3}$, $4.3 \times 10^{-3}$, $6.8 \times 10^{-3}$)} \\
 \hline\hline
  \scriptsize{$M=2\times10^5M_\odot$}&\scriptsize{\text{}} \\ \hline
 \scriptsize{\text{LISA}}&\scriptsize{($3.7 \times 10^{-5}$, $9.6 \times 10^{-4}$, $6.5 \times 10^{-3}$, $2.1 \times 10^{-1}$, $2.1 \times 10^{-1}$)
} \\ \hline
 \scriptsize{\text{Taiji}}&\scriptsize{($2.1 \times 10^{-5}$, $5.6 \times 10^{-4}$, $2.5 \times 10^{-3}$, $6.3 \times 10^{-2}$, $6.3 \times 10^{-2}$)} \\ \hline
 \scriptsize{\text{TianQin}}&\scriptsize{($2.0 \times 10^{-3}$, $5.0 \times 10^{-2}$, $2.9 \times 10^{-1}$, $1.3 \times 10^{2}$, $1.3 \times 10^{2}$)} \\ \hline
 \scriptsize{\text{network}}&\scriptsize{($1.8 \times 10^{-5}$, $4.7 \times 10^{-4}$, $1.7 \times 10^{-3}$, $4.5 \times 10^{-2}$, $4.5 \times 10^{-2}$)} \\
 \hline\hline
  \scriptsize{$M=2\times10^6M_\odot$}&\scriptsize{\text{}} \\ \hline
 \scriptsize{\text{LISA}}&\scriptsize{($5.1 \times 10^{-3}$, $9.4 \times 10^{-2}$, $2.8 \times 10^{-1}$, $8.0 \times 10^{1}$, $8.0 \times 10^{1}$)} \\ \hline
 \scriptsize{\text{Taiji}}&\scriptsize{($1.4 \times 10^{-3}$, $2.9 \times 10^{-2}$, $8.1 \times 10^{-2}$, $2.0 \times 10^{1}$, $2.0 \times 10^{1}$)} \\ \hline
 \scriptsize{\text{TianQin}}&\scriptsize{($1.2 \times 10^{-1}$, $2.5 \times 10^{0}$, $6.2 \times 10^{0}$, $1.4 \times 10^{4}$, $1.4 \times 10^{4}$)} \\ \hline
 \scriptsize{\text{network}}&\scriptsize{($1.3 \times 10^{-3}$, $2.7 \times 10^{-2}$, $6.1 \times 10^{-2}$, $1.5 \times 10^{1}$, $1.5 \times 10^{1}$)} \\
 \hline\hline
\end{tabular}
\caption{The errors of five ppE parameters $(\Delta\alpha_D,\Delta\alpha_Q, \Delta\alpha_V, \Delta\alpha_B, \Delta\alpha_L)$ for different detectors and their network with the $(\mathcal{A},\mathcal{E}, \mathcal{T})$ combination.
The results shown are the median values, obtained by simulating 100 GW sources from different MBHBs with randomly sampled parameters $(\theta_s,\phi_s, \psi,\iota)$, 
where $\cos\theta_s \sim U\left(-1, 1\right)$, $\phi_s\sim U\left(-\pi, \pi\right)$, $\psi_s\sim U\left(0, \pi\right)$ and $\cos\iota \sim U\left(-1, 1\right)$.}
    \label{reserror1}
\end{table}

\begin{table}
    \tabcolsep=1.5cm
    \renewcommand\arraystretch{0.5}
    \begin{tabular}{c | c}\hline\hline
\scriptsize{$M=2\times10^4M_\odot$}&\scriptsize{\text{no-TDI}} \\ \hline
 \scriptsize{\text{LISA}}&\scriptsize{($1.3 \times 10^{-6}$, $9.9 \times 10^{-5}$, $5.3 \times 10^{-3}$, $9.4 \times 10^{-2}$, $9.8 \times 10^{-2}$)} \\ \hline
 \scriptsize{\text{Taiji}}&\scriptsize{($8.9 \times 10^{-7}$, $6.3 \times 10^{-5}$, $2.8 \times 10^{-3}$, $3.1 \times 10^{-2}$, $3.3 \times 10^{-2}$)} \\ \hline
 \scriptsize{\text{TianQin}}&\scriptsize{($3.7 \times 10^{-6}$, $2.5 \times 10^{-4}$, $9.6 \times 10^{-3}$, $1.3 \times 10^{0}$, $1.3 \times 10^{0}$)} \\ \hline
 \scriptsize{\text{network}}&\scriptsize{($6.8 \times 10^{-7}$, $4.9 \times 10^{-5}$, $2.0 \times 10^{-3}$, $2.6 \times 10^{-2}$, $2.7 \times 10^{-2}$)} \\
 \hline\hline
  \scriptsize{$M=2\times10^5M_\odot$}&\scriptsize{\text{}} \\ \hline
 \scriptsize{\text{LISA}}&\scriptsize{($6.7 \times 10^{-6}$, $2.4 \times 10^{-4}$, $3.0 \times 10^{-3}$, $1.2 \times 10^{-1}$, $1.2 \times 10^{-1}$)} \\ \hline
 \scriptsize{\text{Taiji}}&\scriptsize{($4.8 \times 10^{-6}$, $1.8 \times 10^{-4}$, $1.8 \times 10^{-3}$, $5.3 \times 10^{-2}$, $5.4 \times 10^{-2}$)} \\ \hline
 \scriptsize{\text{TianQin}}&\scriptsize{($2.6 \times 10^{-5}$, $9.8 \times 10^{-4}$, $9.0 \times 10^{-3}$, $5.0 \times 10^{0}$, $5.0 \times 10^{0}$)} \\ \hline
 \scriptsize{\text{network}}&\scriptsize{($3.7 \times 10^{-6}$, $1.4 \times 10^{-4}$, $1.1 \times 10^{-3}$, $3.6 \times 10^{-2}$, $3.6 \times 10^{-2}$)} \\
 \hline\hline
  \scriptsize{$M=2\times10^6M_\odot$}&\scriptsize{\text{}} \\ \hline
 \scriptsize{\text{LISA}}&\scriptsize{($1.8 \times 10^{-4}$, $4.0 \times 10^{-3}$, $1.5 \times 10^{-2}$, $4.5 \times 10^{0}$, $4.5 \times 10^{0}$)} \\ \hline
 \scriptsize{\text{Taiji}}&\scriptsize{($6.0 \times 10^{-5}$, $1.5 \times 10^{-3}$, $5.0 \times 10^{-3}$, $1.3 \times 10^{0}$, $1.3 \times 10^{0}$)} \\ \hline
 \scriptsize{\text{TianQin}}&\scriptsize{($3.4 \times 10^{-4}$, $8.3 \times 10^{-3}$, $2.2 \times 10^{-2}$, $1.1 \times 10^{2}$, $1.1 \times 10^{2}$)} \\ \hline
 \scriptsize{\text{network}}&\scriptsize{($5.4 \times 10^{-5}$, $1.3 \times 10^{-3}$, $3.5 \times 10^{-3}$, $1.0 \times 10^{0}$, $1.0 \times 10^{0}$)} \\
 \hline\hline
\end{tabular}
\caption{The errors of five ppE parameters $(\Delta\alpha_D,\Delta\alpha_Q, \Delta\alpha_V, \Delta\alpha_B, \Delta\alpha_L)$ for different detectors and their network without TDI.
The results shown are the median values, obtained by simulating 100 GW sources from different MBHBs with randomly sampled parameters $(\theta_s,\phi_s, \psi,\iota)$, 
where $\cos\theta_s \sim U\left(-1, 1\right)$, $\phi_s\sim U\left(-\pi, \pi\right)$, $\psi_s\sim U\left(0, \pi\right)$ and $\cos\iota \sim U\left(-1, 1\right)$.}
    \label{reserror2}
\end{table}
The mean values of the errors of five ppE parameters $(\Delta\alpha_D,\Delta\alpha_Q, \Delta\alpha_V, \Delta\alpha_B, \Delta\alpha_L)$ for different detectors with and without TDI are shown in Table \ref{reserror1} and Table \ref{reserror2}.
In Fig. \ref{ratiores}, we show the mean and the 1$\sigma$ values of the ratio between the errors of five ppE parameters for LISA, Taiji, TianQin, and their network with and without TDI combination, as well as the mean values of the ratio between the SNR.
The mean values of errors in five ppE parameters and SNR for different detectors with the $(\mathcal{A},\mathcal{E}, \mathcal{T})$ combination are shown in Fig. \ref{TDIres}.


\subsection{The effects of TDI on polarization constraint}
\begin{figure}
  \centering
	\subfloat{
	\includegraphics[width=0.98\textwidth]{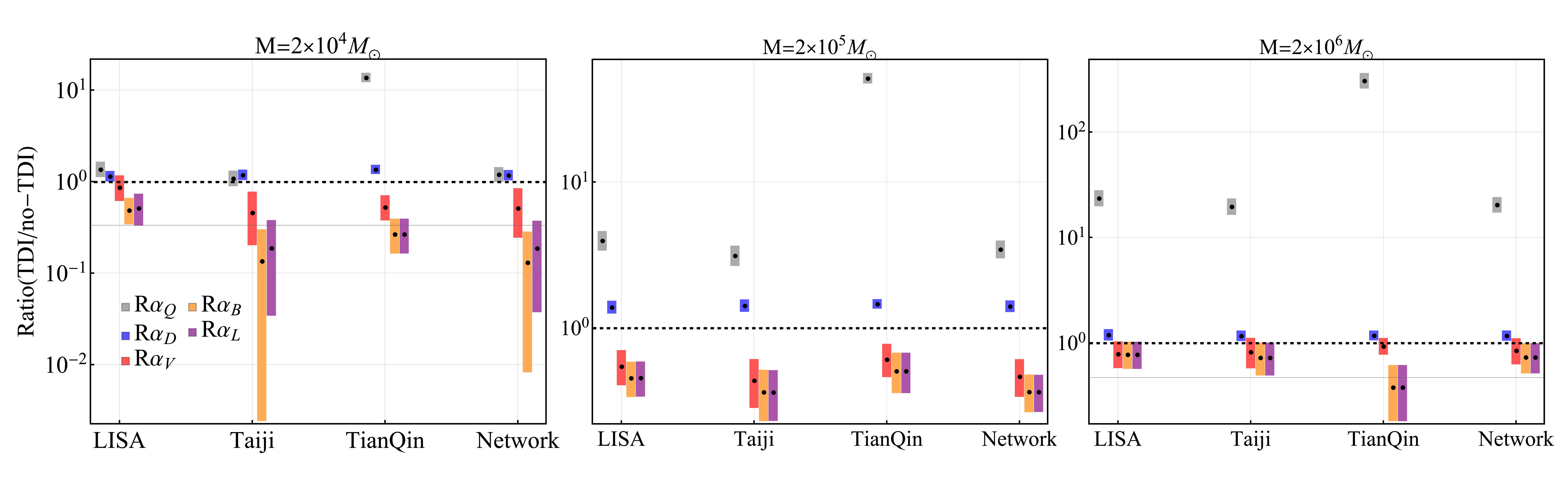}}\\
	  \subfloat{
      \includegraphics[width=0.98\textwidth]{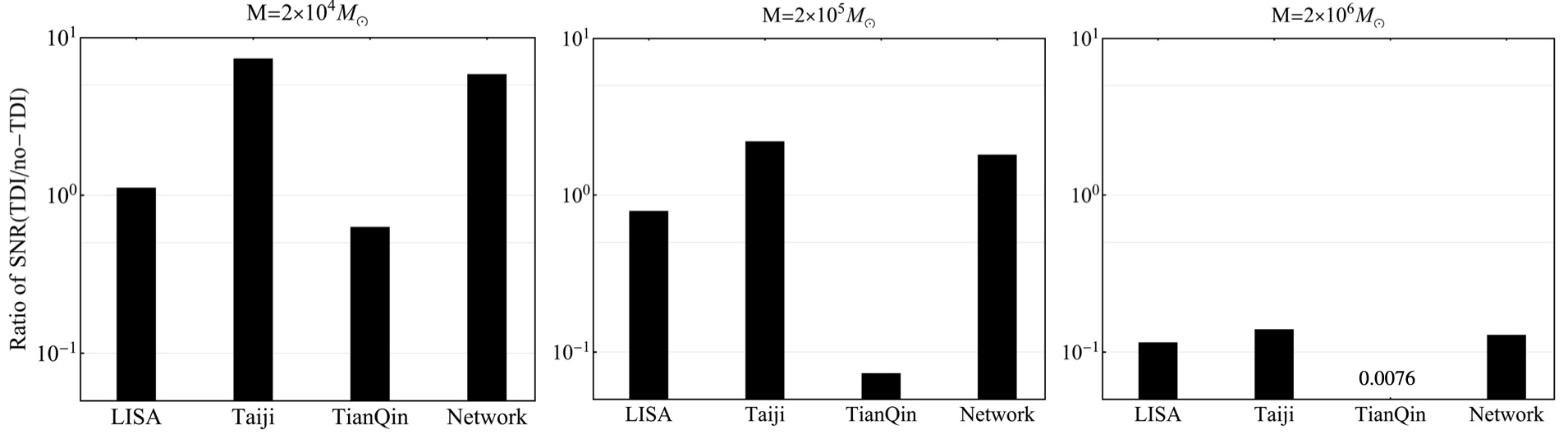}}
\caption{The above panels are the mean and the 1$\sigma$ values of the ratio $R_{\alpha_i}$ between the errors in five ppE parameters for different detectors with and without TDI combination, the colored boxes and the points in the boxes represent the 1$\sigma$ range and the mean values of $R_{\alpha_i}$.
The horizontal dashed line in the panels represents that the ratio equals to 1.
The bottom panels are the mean values of the SNR ratio between the detectors with and without the TDI combination.
The panels from left to right represent the GW sources with $M=(2\times10^4,~2\times10^5,~2\times10^6) M_\odot$.
}
  \label{ratiores}
\end{figure}

To study the impact of TDI on the constraint of extra polarization, we take the optimal combination $(\mathcal{A},\mathcal{E}, \mathcal{T})$ in signal analysis, and give the errors of five ppE parameters $(\Delta\alpha_D,\Delta\alpha_Q, \Delta\alpha_V, \Delta\alpha_B, \Delta\alpha_L)$ for LISA, Taiji, TianQin, and their network with and without TDI combination.
The mean values of the errors are shown in Table \ref{reserror1} and \ref{reserror2}.
The fiducial values we choose $\alpha_D=0$, $\alpha_Q=96/5$, $\alpha_V=0$, $\alpha_L=0$, $\alpha_B=0$, represent GW radiation with only quadrupole radiation in GR, and the error values on the extra polarization parameters $\Delta\alpha_i$ reflect the degree of deviation from GR.
Therefore, we can take the relative error of tensor modes $\Delta\alpha_Q/\alpha_Q$ as the benchmark, and define the factor $\mathcal{F}_i=\frac{\Delta\alpha_i}{\Delta\alpha_Q/\alpha_Q}$ to represent the detector's constraint ability on the extra polarizations.
We use $R_{\alpha_i}$, which is the ratio of the factor $\mathcal{F}_i$ between the results with and without TDI, to study the impact of TDI on the constraint of extra polarization for different detectors.
For the tensor modes, the ratio is $\Delta\alpha_Q^{\rm{TDI}}/\Delta\alpha_Q^{\rm{no-TDI}}$.
The results are shown in Fig. \ref{ratiores}, along with the mean values and 1$\sigma$ uncertainty ranges.
The larger the factors $R_{\alpha_i}$, the worse the constraint ability on the parameter $\alpha_i$ for the detectors with TDI.

Considering the impact of TDI on constraint of tensor modes, as shown in Fig. \ref{ratiores}, we compare the ratio values $R_{\alpha_Q}$ for different detectors.
At low frequencies, when the detector uses the TDI method, the SNR obtained is lower than that using the simplified equal-arm Michelson interferometer method due to signal cancellation effects.
The values of $R_{\alpha_Q}$ for the detectors are all larger than 1, and increase with the more massive GW sources.
This indicates that the constraint on the tensor mode gets worse with TDI method.
Because the GW signal is dominated by quadrupole radiation in the GR case with the fiducial parameters $\alpha_D=0$, $\alpha_Q=96/5$, $\alpha_V=0$, $\alpha_L=0$, $\alpha_B=0$, and the impact of the constraint on tensor mode is entirely dependent on the size of the values of SNR.
When $M>10^5~M_\odot$, the mass of GW source increases, the values of SNR decrease, and this causes the constraint on the tensor mode to get worse for different detectors with the TDI method.
For LISA, the values of $R_{\alpha_Q}$ are $(1.36,~3.97,~23.58)$ for GW sources with $M=(2\times10^4,~2\times10^5,~2\times10^6) M_\odot$.
For Taiji, the values of $R_{\alpha_Q}$ are $(20.4\%,~21\%,~16.7\%)$ less than LISA, and the constraint on tensor mode is slightly better.
Due to the detector configuration being the same with LISA but with longer arm length, it has better sensitivity.
For TianQin, because it is sensitive to GWs in the high frequency, it has the best constraint on tensor mode for the GW source with $M=2\times10^4 M_\odot$, while the mass increases, the SNR rapidly decreases.
The values of $R_{\alpha_Q}$ are $(13.73,~51.42,~306.9)$

The ppE parameters $(\alpha_V, \alpha_B, \alpha_L)$ correspond to the vector mode, scalar breathing mode, and scalar longitudinal mode, respectively.
The parameter $\alpha_D$ characterizes the dipole radiation component in GWs, which dominates in the extra polarization modes.
For these parameters, the definition of factor $R_{\alpha_i}$ has eliminated the influence of variations in the SNR.
Therefore, as shown in Fig. \ref{ratiores}, the value of $R_{\alpha_i}$ for different GW sources are all within the same order of magnitude for the same detector, differing by at most several times, but the $R_{\alpha_D}$ is slightly greater than 1, $(R_{\alpha_V}, R_{\alpha_B},R_{\alpha_L})$ are all less than 1.
This is because while calculating parameter errors from the Fisher matrix, the derivatives of the GW signal with respect to the $\alpha_D$ include the contribution of the phase function Eq. \eqref{phit}, however, the results for the $(\alpha_V, \alpha_B, \alpha_L)$ only contain the amplitude term.
It can be observed that for the parameter $\alpha_D$, its formula of Fisher matrix coincides with the tensor mode; whereas for the parameters $(\alpha_V, \alpha_B, \alpha_L)$, their calculation depends on the specific modified gravity theory.
Consequently, the values of $R_{\alpha_{V, B, L}}$ decrease, and the constraint on extra polarization gets better for different detectors with TDI method.

For the scalar breathing mode and scalar longitudinal mode, their signal formula are the same as the ppE waveform.
Therefore, the values of $R_{\alpha_{B}}$ and $R_{\alpha_{L}}$ remains nearly identical for different detectors.
For LISA, the values of $R_{\alpha_{B/L}}$ are $(0.49,~0.45,~0.78)$ for different GW sources;
For Taiji, the values of $R_{\alpha_{B/L}}$ are $(67.7\%,~19.9\%,~7.1\%)$ less than LISA;
For TianQin, the values of $R_{\alpha_{B/L}}$ are $(0.27,~0.51,~0.38)$.

For the vector mode, its ppE waveform exhibits dependence on inclination angle.
Comparing with the scalar mode, therefore, the constraint on the $\alpha_V$ is affected by the inclination angle.
The angle-averaged results in Fig. \ref{ratiores} show that the value of $R_{\alpha_{V}}$ is slightly larger than $R_{\alpha_{B/L}}$.
For LISA, the values of $R_{\alpha_{V}}$ are $(0.87,~0.54,~0.78)$ for different GW sources;
For Taiji, the values of $R_{\alpha_{V}}$ are $(47\%,~19.9\%,~4.6\%)$ less than LISA;
For TianQin, the values of $R_{\alpha_{V}}$ are $(0.53,~0.61,~0.93)$.

For the network of combined detectors, the results of the impact of TDI on extra polarizations constraints are the same as the single detectors.

\subsection{Constraint on polarization for different detectors with TDI}
\begin{figure}
	\centering
	\subfloat{
	\includegraphics[width=0.98\textwidth]{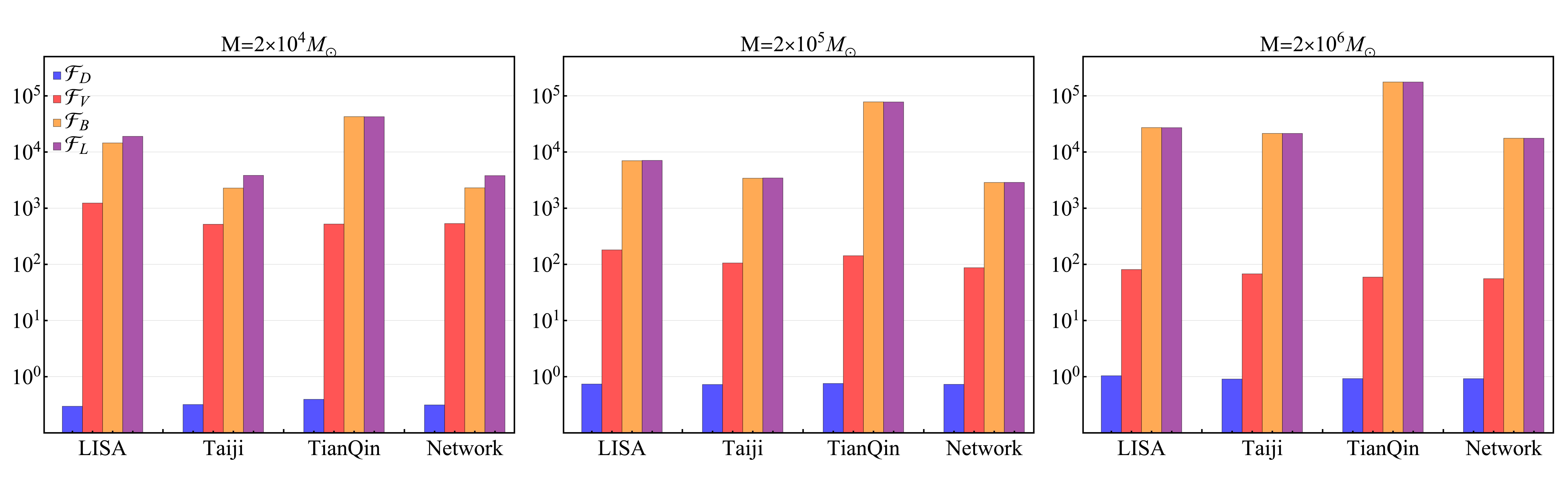}}\\
	\subfloat{
	\includegraphics[width=0.98\textwidth]{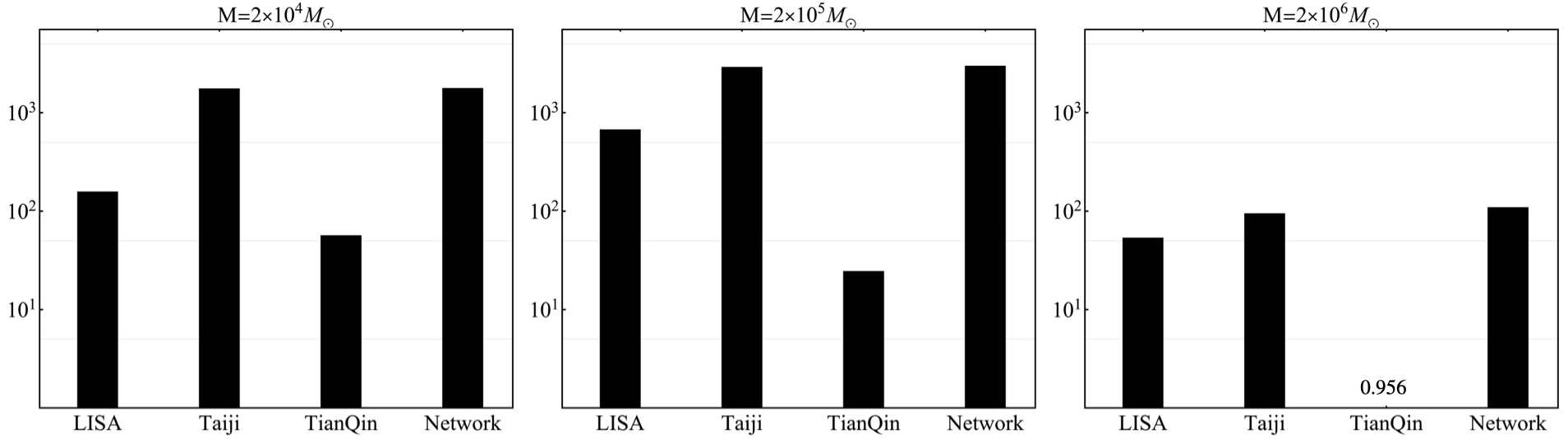}}
\caption{The above panels are mean values of the ratio $\mathcal{F}_i$ factors on the extra polarization parameters for different detectors with $(\mathcal{A},\mathcal{E}, \mathcal{T})$ combination.
The bottom panels are the mean values of the SNR.
The panels from left to right represent the GW sources with total mass $M=(2\times10^4,~2\times10^5,~2\times10^6) M_\odot$.}
  \label{TDIres}
\end{figure}

In this subsection, we apply the second-generation TDI $(\mathcal{A},\mathcal{E}, \mathcal{T})$ combination to study the effects of constraint on polarization for different space-based detectors.
We use the factors $\mathcal{F}_i$ of the ppE parameters $(\Delta\alpha_D,\Delta\alpha_Q, \Delta\alpha_V, \Delta\alpha_B, \Delta\alpha_L)$, to represent the detector's constraint ability on the extra polarizations for different detectors.
The larger the factors $\mathcal{F}_i$, the worse the constraint ability on the parameter $\alpha_i$ for the detectors.
We show the mean values of $\mathcal{F}_i$ and SNR in Fig. \ref{TDIres}.

As shown in Fig. \ref{TDIres}, for the parameter $\alpha_D$, the detector's constraints weaken with increasing total mass.
Because $\alpha_D$ characterizes the dipole radiation in GWs, comparing with the quadrupole radiation, the dipolar effect dominates in a weak-field or low-velocity regime.
So it leads to better constraints at lower masses.
For LISA, the values of $\mathcal{F}_D$  for GW sources with $M=(2\times10^4,~2\times10^5,~2\times10^6) M_\odot$ are $(0.3,~0.74,~1.04)$, while Taiji and TianQin yield nearly identical results, and the difference is less than $33\%$.

Regarding detectors' ability to constrain extra polarizations, the values of $\mathcal{F}_{V,B,L}$ far exceed 1; the results indicate that the detectability of vector, scalar breathing, and scalar longitudinal modes is substantially weaker compared to the tensor mode for different detectors with TDI.
However, for the vector mode, the values of $\mathcal{F}_{V}$ decrease with increasing GW source mass.
For LISA/Taiji/TianQin, the values of $\mathcal{F}_{V}$ are $(1230.47,~180.7,~80.82)$ / $(517.17,~105.65,~67.63)$ / $(522.08,~142.22,~59.12)$ for the three GW sources, respectively.
The detector's constraint ability on the vector mode increases with the more massive GW sources.
The scalar breathing and longitudinal modes exhibit the poorest detectability, with values of $\mathcal{F}_{B/L}$ 1-3 orders of magnitude larger than those of the vector mode.
Nevertheless, due to identical waveform signatures, the detector's constraint ability for them is identical.
For the network, the improvement of the detector's constraint ability on extra polarization is little, due to the detector in different orbital positions can be thought of as an independent detector, the improvement by the network of combined detectors is small.

\section{Discussion}\label{discussion}
In GR, GWs only contain the tensor degrees of freedom, but generic metric theories of gravity may permit up to six independent polarizations.
Therefore, probing the polarization content of GWs with space-based detectors provides the most direct approach to testing the theories of gravity.
Nevertheless, space-based GW detectors cannot maintain fixed interferometric arm lengths due to gravitational free-fall motion, so the TDI method should be considered in signal response.

In this paper, we study the constraint ability of space-based GW detectors on extra polarizations with the second-generation TDI method, and explore the impacts of TDI on the constraint of polarizations.
We adopt the GW waveforms with six degrees of freedom characterized by parameters $(\alpha_D,\alpha_Q, \alpha_V, \alpha_B, \alpha_L)$ in the ppE framework.
Considering the second-generation TDI $(\mathcal{A},\mathcal{E}, \mathcal{T})$ combination, we give the parameter errors with FIM method.
For the fiducial parameter values $\alpha_D=0$, $\alpha_Q=96/5$, $\alpha_V=0$, $\alpha_L=0$, $\alpha_B=0$, the size of values of parameter errors quantifies the degree of deviation from GR.
We subsequently compare the constraint abilities on extra polarization for LISA, Taiji, TianQin and their network with GW sources mass $M=(2\times10^4,~2\times10^5,~2\times10^6) M_\odot$, separately.

For the tensor mode, the detector's constraint ability depends entirely on the SNR.
The SNR of GWs decreases by employing the TDI method; the reduction is particularly pronounced for more massive GW sources.
This is because at low frequencies, when the detector uses the second-generation TDI method, the SNR obtained is lower than that using the simplified equal-arm Michelson interferometer method due to signal cancellation effects.
Therefore, it degrades the constraints on the tensor mode with TDI method.
Comparing the different detectors, constraint ability for the tensor mode is related to the sensitivity.
Thus, Taiji achieves marginally stronger constraints than LISA, while TianQin exhibits the weakest results.

For extra polarizations, the GWs are primarily dominated by dipole radiation.
When the sources reside in weak-field or low-velocity regimes, the dipole radiation intensifies, resulting in enhanced constraints for the parameter $\alpha_D$ at lower masses.
However, the results show that the impact TDI on constraint of parameter $\alpha_D$ is negligible, while detectors' constraint ability for extra polarizations is fundamentally governed by the theory of gravity.
For the scalar breathing and longitudinal modes, their identical waveform in ppE framework leads to identical constraint variations with TDI method.
Both modes exhibit marginal improvement compared to no-TDI scenarios.
For the vector mode, the constraint depends on the inclination angle, and it yields a slight enhancement for the detectors' ability with TDI.

Moreover, we compare the detection's ability on extra polarizations for different detectors with TDI $(\mathcal{A},\mathcal{E}, \mathcal{T})$ combination.
Comparing with the tensor mode, the detector's ability on vector, scalar breathing, and scalar longitudinal modes are substantially weaker.
However, for the vector mode, detectability improves with increasing gravitational wave source mass.
The scalar breathing and longitudinal modes exhibit the poorest detectability, with errors exceeding those of the vector mode by 1-3 orders of magnitude. Due to their identical waveform signatures in ppE framework, however, their detectability are the same.
For the network, the improvement of the detector's constraint ability on extra polarization is little, because the detector in different orbital positions can be thought of as an independent detector, which causes the weak effect for combined detectors.

\begin{acknowledgments}
This research is supported by the National Natural Science Foundation of China (NSFC) under Grant No. 12465013.
\end{acknowledgments}

%

\end{document}